\documentclass[12pt]{iopart}

\usepackage{iopams}  
\usepackage{graphicx}

\begin{document}
\title {Dynamics of a driven spin coupled to an antiferromagnetic
spin bath}

\author{Xiao-Zhong Yuan$^{1,2}$, Hsi-Sheng Goan$^{1,3}$\footnote{Author to whom any correspondence should be addressed.} and Ka-Di Zhu$^2$}

\address{$^1$ Department of Physics and Center for Theoretical Sciences,
National Taiwan University, Taipei 10617, Taiwan}
\address{$^2$ Department of Physics, Shanghai Jiao Tong University,
Shanghai 200240, China}
\address{$^3$ Center for Quantum Science and Engineering,
National Taiwan University, Taipei 10617, Taiwan}

\ead{goan@phys.ntu.edu.tw}

\begin{abstract}
We study the behavior of the Rabi oscillations of a driven central spin 
(qubit) coupled to an antiferromagnetic spin bath (environment). 
It is found that the decoherence
behavior of the central spin depends on the
detuning, driving strength, the qubit-bath coupling 
and an important factor, associated with the number of the
coupled atoms, 
the detailed lattice structure, and the temperature of the
environment. If the detuning exists, the Rabi oscillations may
show the behavior of collapses and revivals; however, if the
detuning is zero, such a behavior will not appear. We investigate
the weighted frequency distribution of the time evolution of the
central spin inversion and give this
phenomenon of collapses and revivals a reasonable explanation. 
We also discuss the
decoherence and the pointer states of the qubit from the
perspectives of the von Neumann entropy.
It is found that the eigenstates of the qubit self-Hamiltonian emerge
as the pointer states in the weak system-environment coupling limit.
\end{abstract}


\submitto{\NJP}

\maketitle

\section{Introduction}
One of the most promising candidates for quantum computation is
the implementation using
spin systems as quantum bits (qubits)
\cite{Kane,Goan05,Hill03,Goan10,Burkard,Raussendorf,Hu}. Combining with
nanostructure technology, they have the potential advantage of
being scalable to a large system. In particular, the
spin in a quantum dot exhibits a relatively long coherence time
compared with fast 
gate operation times. This makes it a good candidate for a quantum
information carrier \cite{loss,Cerletti,Borhani,Tackeuchi}.
However the influence of the environment, especially the spin
environment, on a spin, which usually causes
the decoherence of the spin, is inevitable. Typically, the material
environment must be present in order to host the spin or to
locally control the electric or magnetic fields experienced by the
spin. Therefore the decoherence behavior of a central spin or
several spins interacting with a spin bath has attracted
much attention in recent years
\cite{Cywinski,WXZhang,WXZhang2,WXZhang3,JLages,Dobrovitski,Takahashi,Yang,Yang2,Witzel9,Dobrovitski3,Hassanieh,Cucchietti,Dziarmaga}.

The problem of a central spin coupled to an antiferromagnetic (AF) spin
environment was investigated in Refs.~\cite{Yuanepl,YuanNJP,Yuan10}.
In these investigations, a
pure dephasing model in which the self-Hamiltonian of the
undriven central spin commutes
with the interaction Hamiltonian to the environment was considered.
A similar problem of
a spin-$\frac{1}{2}$ impurity embedded in an AF environment was
studied in the context of quantum frustration of decoherence 
in Ref.~\cite{Novais05}. There the impurity spin is coupled locally in
real space to just one spin of the AF environment, in
contrast to the central spin model \cite{Breuer04a,Bose04,Lucamarini04,Breuer04b,Hamdouni06,Yuan07,Quan06,Cucchietti07,Rossini07,Lidar07,Breuer08,Lidar10}
where the central spin is coupled isotropically 
with equal strength to all the spins
of the environment.
In this paper, 
we consider a more general case of a central
spin (qubit) driven by an external field
and coupled to an AF spin bath (environment). 
To control the quantum states
of a qubit, some types of  
time-dependent controllable manipulations 
, which can be electrical \cite{Pettaa}, optical
\cite{Mikkelsenn}, or 
magnetic \cite{Koppenss}, are desirable. 
Thus it is important and necessary
to study the behavior of
a qubit (central spin) or several qubits (spins)
interacting with an environment (a spin bath) 
in the presence of a driving field
\cite{131,Koppenss2,Hanson,Rao}. Recently, the coherent dynamics
of a single central spin (a nitrogen-vacancy center) coupled to a
bath of spins (nitrogen impurities) in diamond was studied
experimentally \cite{Hanson}. To realize the manipulation of the
central spin, the pulsed radio-frequency radiation was used. 
Under the control of a time-dependent magnetic field,
Ref.~\cite{Rao} considered a central spin system coupled to a spin
bath. The decay of Rabi oscillations and the loss of entanglement
were discussed. However, the correlations between spins in the spin bath were
neglected. Recent investigations indicated that the internal dynamics
of the spin bath could be crucial to the decoherence of the 
central spin
\cite{Cywinski,WXZhang,WXZhang2,WXZhang3,JLages,Dobrovitski,Takahashi,Yang,Yang2,Witzel9,Dobrovitski3}.
In our driven spin model, the interactions between constituent spins of
the AF environment are taken into account.

The self-Hamiltonian of the central spin, after a
transformation to a frame rotating with the frequency of the driving field, 
can be written in a form of  
$H_S=\varepsilon S_0^z+gS_0^x$ in the rotating wave approximation.
Due to the additional driving term of $g S_0^x$ that provides energy into the
system and 
does not commute with the interaction Hamiltonian that couples to the AF
environment in our model, the dynamics of the central 
spin in this case is
dramatically different from the undriven 
pure dephasing behaviors investigated in
Refs.~\cite{Yuanepl,YuanNJP}. 
It was shown in Ref.~\cite{Dobrovitski} that the form
and the rate of Rabi oscillation decay are useful in experimentally
determining the intrabath coupling strength for a broad class of
solid-state systems \cite{Dobrovitski}. This is also the case in
our problem.
After the use of the spin-wave
approximation to deal with the AF environment,
we employed an elegant mathematic technique to
trace over the AF environmental degrees of freedom exactly
to obtain  the reduced density 
matrix of the driven spin.
This enables us
to study and describe the decay behaviors of the Rabi oscillations for different initial states and
parameters of the central spin-AF environment model beyond the
Markovian approximation and Born approximation (perturbation). 
With the reduced density matrix obtained nonperturbatively, 
we also investigate and discuss the
decoherence and the pointer states of the central spin from the
perspective of the von Neumann entropy.
We find that the decoherence
behavior of the central spin depends on the
detuning, driving strength, the coupling between the central spin and the spin
environment, and an important factor $\Omega$, 
associated with the number of the spins, 
the lattice structure and the temperature of the environment.
If the detuning exists, the Rabi oscillations may
show a behavior of collapses and revivals; however, if the
detuning is zero, such a behavior will not appear. We investigate
the weighted frequency distribution of the time evolution of the
central spin inversion and give this
phenomenon a reasonable explanation. 
The form and the rate of Rabi oscillation decay is useful in
determining the intrabath coupling strength and 
other related properties of the qubit-environment system \cite{Dobrovitski}. 
Although we concentrate on the central spin model, our study 
is applicable to similar models of 
a pseudo spin or a qubit coupled to an environment.
For example, in the study of 
the decoherence behavior of a flux qubit interacting with
a spin environment, the same self-Hamiltonian of the qubit,
$H_S=\varepsilon S_0^z+gS_0^x$, can be used \cite{Dziarmaga}. 
In such case, $\varepsilon$ is the bias energy
and $g$ is tunnel splitting of the flux qubit, 
and the two eigenstates of $S_0^z$
correspond to macroscopically distinct states that have a
clockwise or an anticlockwise circulating current \cite{Dziarmaga} which can be
denoted as $|1\rangle$ or $|0\rangle$. 
To make contact of the driven
central spin model with 
experiments, 
one may envisage a setup of  
a small ring-shaped flux qubit located at a distance 
above or below an also ring-shaped
AF material that has a common
symmetric z-axis with the flux qubit. 
In this case, the coupling strength between the
flux qubit and each of the constituent spins of the 
AF material (environment) may be regarded to be almost the same.

The paper is organized as follows. In Sec.~\ref{sec:model}, the model
Hamiltonian is introduced and the spin wave approximation is
applied to map the spin operators of the AF
environment to bosonic operators. After tracing over the
environmental modes, the reduced density matrix is obtained and
the dynamics of the central spin is calculated. 
We also investigate the decoherence and the pointer states 
of the central spin for the cases of
zero detuning and nonzero detuning, and calculate the 
von Neumann entropy which is a measure of the purity of the
mixed state. 
Numerical results and discussions are presented in Sec.~\ref{sec:results}. 
Conclusions are given in Sec.~\ref{sec:conclusions}.

\section{Model and Calculations}
\label{sec:model} 
\subsection{Model and transformed Hamiltonian}
We consider a central spin driven by an external microwave
magnetic field and embedded in
an AF material. To detect the central spin and
control its states, the frequency $\omega_c$ of the microwave
magnetic field is tuned to be resonant or near resonant with the
central spin.  Furthermore, the central spin and the
AF environment are made of spin-$\frac{1}{2}$
atoms. 
The total Hamiltonian of our model can be written as
\begin{eqnarray}
H&=&H_S+H_{SB}+H_B,
\end{eqnarray}
where $H_S$, $H_B$ are the Hamiltonians of the central spin and
the AF environment respectively, 
and $H_{SB}$ is the interaction between them
\cite{Yuanepl,YuanNJP,Lucamarini,Driessen}.
They can be written as ($\hbar=1$)
\begin{eqnarray}
H_S&=&\mu_0S_0^z+g(S_0^+e^{-i\omega_ct}+S_0^-e^{i\omega_ct}),\\
H_{SB}&=&-\frac{J'_0}{\sqrt{N}} S_0^z \sum_{i}
(S_{a,i}^z+S_{b,i}^z),\label{eqpp2}\\
H_B&=&J\sum_{i,\vec{\delta}}\mathbf{S}_{a,i}\cdot\mathbf{S}_{b,i+\vec{\delta}}
+J\sum_{j,\vec{\delta}}\mathbf{S}_{b,j}\cdot\mathbf{S}_{a,j+\vec{\delta}}
\,,
\end{eqnarray}
where $\mu_0$ is the Larmor frequency and 
represents the coupling constant with a local
magnetic field in the $z$ direction. The magnetic field creates a
local Zeemann splitting, which can then be accessed by the driving
field with frequency $\omega_c$ and (real) coupling strength $g$.
The coupling strength $g$ is proportional to 
the amplitude of the driving field.  
We assume that the spin structure of the AF environment may
be divided into two interpenetrating sublattices $a$ and $b$ with
the property that all nearest neighbors of an atom on $a$ lie on
$b$, and \textit{vice versa} \cite{Kittel}. $\mathbf{S}_{a,i}$
($\mathbf{S}_{b,j}$) represents the spin operator of the $i$th
($j$th) atom on sublattice $a$ ($b$). 
The indices $i$ and $j$ label the $N$ atoms in each sublattice, whereas
the vectors $\vec{\delta}$ connect atom $i$ or $j$ with its
nearest neighbors. $J$ is the exchange
interaction and is positive for AF environment.
The effects of the next nearest-neighbor
interactions are neglected.
For simplicity, significant interaction between the
central spin and the environment is assumed to be of the Ising type. 
This type of interaction has gained additional
importance because of its relevance to quantum information
processing \cite{Cucchietti,Koppenss2,Rao,Breuer04}.
The   
coupling constant between the central spin and AF   
environment is scaled as $J'_0/\sqrt{N}$ such that nontrivial   
finite limit of $N\rightarrow\infty$ can
exist \cite{Yuanepl,YuanNJP,Lucamarini,Breuer04,Frasca04}.
For convenience, we denote in the following 
the scaled interaction between the central
spin and the AF environment as $J_0=J'_0/\sqrt{N}$.


In a frame rotating with the frequency of the driving field, the
Hamiltonian of the central spin becomes
\begin{eqnarray}
H_S&=&\varepsilon S_0^z+gS_0^x,
\label{Hs_rot}
\end{eqnarray}
where the detuning 
\begin{equation}
  \label{eq:detuning}
\varepsilon=\mu_0-\omega_c.  
\end{equation}
Using the Holstein-Primakoff transformation,
\begin{eqnarray}
S_{a,i}^+&=&\sqrt{1-a_i^+a_i}a_i,
\hspace*{2mm}S_{a,i}^-=a_i^+\sqrt{1-a_i^+a_i},
\hspace*{2mm}S_{a,i}^z=\frac{1}{2}-a_i^+a_i,
\label{spina}\\
S_{b,j}^+&=&b_j^+\sqrt{1-b_j^+b_j}, \hspace*{2mm}
S_{b,j}^-=\sqrt{1-b_j^+b_j}b_j,
\hspace*{3mm}S_{b,j}^z=b_j^+b_j-\frac{1}{2},
\label{spinb}
\end{eqnarray}
we map spin operators of
the AF environment onto bosonic operators.  
We will consider the situation that the environment is
in the low-temperature and low-excitation limit such that
the spin operators in Eqs.~(\ref{spina}) and (\ref{spinb}) 
can be approximated as
$S_{a,i}^+\approx a_i$, and
$S_{b,j}^+\approx b_j^+$.
This can be justified because in this limit,
the number of excitation is small, and the
thermal averages $<a_i^+a_i>$ and $<b_i^+b_i>$ are expected to be of
the order $O(1/N)$ and can be safely neglected when $N$ is very large.
The Hamiltonians $H_{SB}$ and $H_B$ can then be written in the
spin-wave approximation \cite{Kittel} as 
\begin{eqnarray}
H_{SB}&=&-J_0S_0^z \sum_{i}(b_i^+b_i-a_i^+a_i),
\label{HSB2}\\
H_B&=&-\frac{1}{2}NMJ+MJ\sum_{i}
(a_i^+a_i+b_i^+b_i)+J\sum_{i,\vec{\delta}}(a_ib_{i+\vec{\delta}}+a_i^+b_{i+\vec{\delta}}^+),
\label{HB2}
\end{eqnarray}
where $M$ is the number of the nearest neighbors of an atom.
We note here that in obtaining Hamiltonian (\ref{HB2})
in line with the approximations of 
$S_{a,i}^+\approx a_i$, and
$S_{b,j}^+\approx b_j^+$ in the low excitation limit, 
we have neglected terms that contain
products of four operators. 
The low excitations correspond to
low temperatures, $T\ll T_N$, where $T_N$ is the N\'{e}el
temperature \cite{Yosida}.
Then transforming Eqs.~(\ref{HSB2})and (\ref{HB2}) to the momentum
space, we have 
\begin{eqnarray}
H_{SB}&=&-J_0S_0^z \sum_{\mathbf{k}}(b_\mathbf{k}^+b_\mathbf{k}-a_\mathbf{k}^+a_\mathbf{k}),\\
H_B&=&-\frac{1}{2}NMJ+MJ\sum_{\mathbf{k}}
(a_\mathbf{k}^+a_\mathbf{k}+b_\mathbf{k}^+b_\mathbf{k})\nonumber\\
&&+MJ\sum_{\mathbf{k}}\gamma_{\mathbf{k}}(a_\mathbf{k}^+b_{\mathbf{k}}^++a_\mathbf{k}b_{\mathbf{k}}),
\end{eqnarray}
where
$\gamma_{\mathbf{k}}=M^{-1}\sum_{\mathbf{\vec{\delta}}}e^{i\mathbf{k}\cdot\mathbf{\vec{\delta}}}$.
Furthermore, by using the Bogoliubov transformation,
\begin{eqnarray}
\alpha_\mathbf{k}=u_\mathbf{k}a_\mathbf{k}-v_\mathbf{k}b_\mathbf{k}^+,\\
 \beta_\mathbf{k}=u_\mathbf{k}b_\mathbf{k}-v_\mathbf{k}a_\mathbf{k}^+,
\end{eqnarray}
where $u_\mathbf{k}^2=(1+\Delta)/2$,
$v_\mathbf{k}^2=-(1-\Delta)/2$, 
$\Delta=1/\sqrt{1-\gamma_{\mathbf{k}}^2}$, and the Hamiltonians
$H_{SB}$ and $H_B$ can be diagonalized ($\hbar=1$)
and be written as
\begin{eqnarray}
H_{SB}&=&-J_0S_0^z \sum_{\mathbf{k}}
(\beta_\mathbf{k}^+\beta_\mathbf{k}-\alpha_\mathbf{k}^+\alpha_\mathbf{k}),\\
H_B&=&-\frac{3}{2}NMJ+\sum_{\mathbf{k}}\omega_\mathbf{k}\left(\alpha_\mathbf{k}^+\alpha_\mathbf{k}+\beta_\mathbf{k}^+\beta_\mathbf{k}+1\right),
\end{eqnarray}
where $\alpha_\mathbf{k}^+$ ($\alpha_\mathbf{k}$) and
$\beta_\mathbf{k}^+$ ($\beta_\mathbf{k}$) are the creation
(annihilation) operators of the two different magnons with
wavevector ${\mathbf{k}}$ and frequency $\omega_{\mathbf{k}}$
respectively. 
For a cubic crystal system in the small $k$
approximation,
\begin{eqnarray}
\omega_\mathbf{k}&=&(2M)^{1/2}Jkl,
\end{eqnarray}
where $l$ is the side length of cubic primitive cell of the
sublattice. 
The constants $-\frac{3}{2}NMJ$ and
$\sum_{\mathbf{k}}\omega_\mathbf{k}$ can be neglected. 
Finally, the
transformed Hamiltonians become
\begin{eqnarray}
H_S&=&\varepsilon S_0^z+g S_0^x,\label{HSf}\\
H_{SB}&=&-J_0S_0^z \sum_{\mathbf{k}}
(\beta_\mathbf{k}^+\beta_\mathbf{k}-\alpha_\mathbf{k}^+\alpha_\mathbf{k}),
\label{HSBf}\\
H_B&=&\sum_{\mathbf{k}}\omega_\mathbf{k}\left(\alpha_\mathbf{k}^+\alpha_\mathbf{k}+\beta_\mathbf{k}^+\beta_\mathbf{k}\right).
\label{HBf}
\end{eqnarray}
Similar to the famous spin-boson model \cite{Leggett87,Irish}, the interaction
Hamiltonian does not commute with the self-Hamiltonian of the spin. 
However, a significant difference from the spin-boson model
\cite{Leggett87,Irish}
is that 
the interaction Hamiltonian commutes with the bath
Hamiltonian, i.e., $[H_{SB},H_B]=0$. So the problem of the total transformed
Hamiltonian, Eqs.~(\ref{HSf})--(\ref{HBf}), can be
solved exactly, even in the case of multi-environment modes and finite
environment temperatures.

\subsection{Reduced density matrix}
We assume the initial total density matrix of the composed system is
separable, i.e., $\rho(0)=\rho_S(0)\otimes\rho_B$.  The density matrix
of the AF bath satisfies the Boltzmann distribution
$\rho_B=e^{-H_B/T}/Z$, where $Z$ is the partition function and the
Boltzmann constant has been set to one.
If the initial state of the qubit (central spin) is taken as
$\rho_S(0)=|\psi(0)\rangle \langle\psi(0)|$ where
\begin{eqnarray}
|\psi(0)\rangle=\delta|1\rangle+\gamma|0\rangle,\\
|\delta|^2+|\gamma|^2=1,
\end{eqnarray}
then the reduced density matrix 
operator of the qubit can be written as
\begin{eqnarray}
\rho_S(t) &=&\frac{1}{Z}\textrm{tr}_B \left[e^{-iHt}|\psi(0)\rangle
 e^{-H_B/T}\langle\psi(0)|e^{iHt}\right]\nonumber\\
&=&\frac{1}{Z}|\delta|^2\textrm{tr}_B\left[e^{-i(H_S+H_{SB})t}|1\rangle
e^{-H_B/T}\langle1| e^{i(H_S+H_{SB})t}\right]\nonumber\\
&&+\frac{1}{Z}\delta\gamma^*\textrm{tr}_B\left[e^{-i(H_S+H_{SB})t}|1\rangle
e^{-H_B/T}\langle0| e^{i(H_S+H_{SB})t}\right]\nonumber\\
&&+\frac{1}{Z}\delta^*\gamma\textrm{tr}_B\left[e^{-i(H_S+H_{SB})t}|0\rangle
e^{-H_B/T}\langle1| e^{i(H_S+H_{SB})t}\right]\nonumber\\
&&+\frac{1}{Z}|\gamma|^2\textrm{tr}_B\left[e^{-i(H_S+H_{SB})t}|0\rangle
e^{-H_B/T}\langle0| e^{i(H_S+H_{SB})t}\right],
\label{reduced-rhos}
\end{eqnarray}
where $\textrm{tr}_B$ denotes the partial trace taken over the
Hilbert space $H_B$ of the environment.
The partition function $Z$ can be evaluated as 
\begin{eqnarray}
Z&=&\textrm{tr}_Be^{-H_B/T}\nonumber\\
&=&\textrm{tr}_Be^{-\sum_{\mathbf{k}}\omega_\mathbf{k}\left(\alpha_\mathbf{k}^+\alpha_\mathbf{k}+\beta_\mathbf{k}^+\beta_\mathbf{k}\right)/T}\nonumber\\
&=&\left(\prod_{\mathbf{k}}\frac{1}{1-e^{-\omega_\mathbf{k}/T}}\right)^2\nonumber\\
&=&e^{-2\sum_\mathbf{k} \ln \left (1-e^{-\omega_\mathbf{k}/T}
\right)}\nonumber\\
 &=&e^{-2\frac{V}{8\pi^3}\int \ln \left
(1-e^{-\omega_\mathbf{k}/T} \right)4\pi k^2dk},\label{eq2}
\end{eqnarray}
where $V$ is the volume of the environment. 
At a low temperature
such that $\omega_{max}>>T$, we may extend the
upper limit of the integration to infinity.
With $x=(2M)^{1/2}Jkl/T$ and
$N=V/l^3$, we obtain
\begin{eqnarray}
Z&=& e^{-2\Omega\int_0^\infty \ln \left ( 1-e^{-x} \right) x^2dx},
\end{eqnarray}
where
\begin{eqnarray}
\Omega=\frac{NT^3}{4\sqrt{2}\pi^2M^{3/2}J^3} \, .
\label{eq:Omega}
\end{eqnarray}

To obtain the exact density matrix operator, Eq.~(\ref{reduced-rhos}), 
we need to evaluate $e^{-i(H_S+H_{SB})t}|1\rangle$ and  
$e^{-i(H_S+H_{SB})t}|0\rangle$.
To proceed, 
we adopt a special operator technique presented in Ref.~\cite{Yuan07}.
We can see that the Hamiltonian $H_S+H_{SB}$ contains operators
$\alpha_\mathbf{k}^+$ , $\alpha_\mathbf{k}$, $\beta_\mathbf{k}^+$,
$\beta_\mathbf{k}$, $S_0^-$, $S_0^+$, and $S_0^z$, where $S_0^-$
and $S_0^+$ change the system state from $|1\rangle$ to
$|0\rangle$, and \textit{vice versa}. 
It is then obvious that we can write
\begin{eqnarray}
e^{-i(H_S+H_{SB})t}|1\rangle=A|1\rangle+B|0\rangle, \label{eq3}
\end{eqnarray}
where $A$ and $B$ are functions of operators $\alpha_\mathbf{k}^+$
, $\alpha_\mathbf{k}$, $\beta_\mathbf{k}^+$ , $\beta_\mathbf{k}$,
and time $t$. 
Using the Schr\"{o}dinger equation identity
\begin{eqnarray}
i\frac{d}{dt}\left[e^{-i(H_S+H_{SB})t}|1\rangle\right]=(H_S+H_{SB})\left[e^{-i(H_S+H_{SB})t}|1\rangle\right],
\end{eqnarray}
and Eq.~(\ref{eq3}), we obtain
\begin{eqnarray}
i\frac{d}{dt}A&=&\frac{g}{2}B-\frac{1}{2}\left[J_0\sum_{\mathbf{k}}(\beta_\mathbf{k}^+\beta_\mathbf{k}-\alpha_\mathbf{k}^+\alpha_\mathbf{k})-\varepsilon\right]A,\label{eq12}\\
i\frac{d}{dt}B&=&\frac{g}{2}
A+\frac{1}{2}\left[J_0\sum_{\mathbf{k}}(\beta_\mathbf{k}^+\beta_\mathbf{k}-\alpha_\mathbf{k}^+\alpha_\mathbf{k})-\varepsilon\right]B, \label{eq13}
\end{eqnarray}
with the initial conditions from Eq.~(\ref{eq3}) given by
\begin{eqnarray}
A(0)&=&1,\\
B(0)&=&0.
\end{eqnarray}
We note that the coefficients of Eqs.~(\ref{eq12}) and (\ref{eq13})
involve only the operators $\alpha_\mathbf{k}^+\alpha_\mathbf{k}$
and $\beta_\mathbf{k}^+\beta_\mathbf{k}$. As a result, we know
that $A$ and $B$ are functions of
$\alpha_\mathbf{k}^+\alpha_\mathbf{k}$,
$\beta_\mathbf{k}^+\beta_\mathbf{k}$, and $t$. They therefore
commute with each other. Consequently, we can treat 
Eqs.~(\ref{eq12}) and (\ref{eq13}) as 
coupled complex-number differential equations and solve
them in a usual way. 
This operator approach allows us to solve Eq.~(\ref{eq3}) and obtain
\begin{eqnarray}
A&=&\cos(\kappa
t/2)+i\frac{J_{0}\sum_{\mathbf{k}}(\beta_\mathbf{k}^+\beta_\mathbf{k}-\alpha_\mathbf{k}^+\alpha_\mathbf{k})-\varepsilon}{\kappa}\sin(\kappa
t/2),
\label{coeff_A}\\
B&=&-i\frac{g}{\kappa}\sin(\kappa t/2),
\label{coeff_B}
\end{eqnarray}
where
\begin{eqnarray}
\kappa=\sqrt{\left[J_0\sum_{\mathbf{k}}(\beta_\mathbf{k}^+\beta_\mathbf{k}-\alpha_\mathbf{k}^+\alpha_\mathbf{k})-\varepsilon\right]^2+g^2}
\, .
\label{eq6}
\end{eqnarray}
Following the similar calculations above, we can evaluate
the time evolution for the initial spin state of $|0\rangle$.
Let
\begin{eqnarray}
e^{-i(H_S+H_{SB})t}|0\rangle=C|1\rangle+D|0\rangle.
\label{initial_0}
\end{eqnarray}
In a similar way, we obtain
\begin{eqnarray}
C&=&B,
\label{coeff_C}\\
D&=&A^+.
\label{coeff_D}
\end{eqnarray}
With Eqs.~(\ref{eq3}), (\ref{coeff_A}), (\ref{coeff_B}),
(\ref{initial_0}), (\ref{coeff_C}) and (\ref{coeff_D}),
the reduced density matrix operator,  Eq.~(\ref{reduced-rhos}),
can be obtained analytically using a particular mathematical method to deal with
the trace over the degrees of freedom of the thermal AF environment.
This particular mathematical 
method will be described in subsection \ref{sec:dynamics}.

Note that our approach also applies to the case in which
the qubit (central spin) is initially in a mixed state. For example,
if the initial state for the qubit is 
$\rho_S(0)=|\delta|^2 |1\rangle \langle 1|+|\gamma|^2|0\rangle \langle
0|$, the corresponding reduced density
matrix is  Eq.~(\ref{reduced-rhos}) provided that the second and third
terms that contain, respectively, $\delta\gamma^*$ and $\delta^*\gamma$
on its right-hand side are removed.

\subsection{Dynamics of the central spin}
\label{sec:dynamics}
With the time evolution of the 
density matrix operator, 
we can investigate the dynamical behavior of the
central spin, which is of particular interest from the perspective
of practical application. 
Here we discuss the time
dependence of the expectation value of $S_0^z(t)$. It can be written
as
\begin{eqnarray}
\langle S_0^z(t)\rangle&=&\textrm{tr}(S_0^z\rho_S(t))\nonumber\\
&=&\langle1|S_0^z\rho_S(t)|1\rangle+\langle0|S_0^z\rho_S(t)|0\rangle\nonumber\\
&=&\frac{1}{2Z}|\delta|^2\textrm{tr}_B
\left[\left(A^+A-B^+B\right)e^{-H_B/T}\right]\nonumber\\
&&+\frac{1}{2Z}|\gamma|^2\textrm{tr}_B
\left[\left(C^+C-D^+D\right)e^{-H_B/T}\right]\nonumber\\
&&+\frac{1}{2Z}\delta\gamma^*\textrm{tr}_B
\left[\left(AC^+-BD^+\right)e^{-H_B/T}\right]\nonumber\\
&&+\frac{1}{2Z}\delta^*\gamma\textrm{tr}_B
\left[\left(A^+C-B^+D\right)e^{-H_B/T}\right].\label{eq7}
\end{eqnarray}
As demonstrated in Eq.~(\ref{eq2}), by decomposing the
Hamiltonian $H_B$ and tracing over the modes of $\alpha_\mathbf{k}$
and $\beta_\mathbf{k}$ separately, the
partition function $Z$ can be calculated. 
But according to the
expression of Eqs.~(\ref{coeff_A})-(\ref{eq6}), it is impossible to do
so in Eq.~(\ref{eq7}). Here we introduce a particular mathematical
method to perform the trace over the environment degrees of freedom.
Our procedure takes two steps. First,
the states with definite eigenvalues, say $P_1$ and $P_2$, of the 
respective operators
$\sum_{\mathbf{k}}\alpha_\mathbf{k}^+\alpha_\mathbf{k}$ and
$\sum_{\mathbf{k}}\beta_\mathbf{k}^+\beta_\mathbf{k}$ are traced
over. Then 
we sum over all possible eigenvalues $P_1$ and $P_2$ 
of the operators
$\sum_{\mathbf{k}}\alpha_\mathbf{k}^+\alpha_\mathbf{k}$ and
$\sum_{\mathbf{k}}\beta_\mathbf{k}^+\beta_\mathbf{k}$.
In this way, we obtain the final expression of $\langle
S_0^z(t)\rangle$, i.e.,
\begin{eqnarray}
\langle
S_0^z(t)\rangle&=&\frac{1}{2Z}\sum_{P_1=0}^\infty\sum_{P_2=0}^\infty
\left[f_1(P_1,P_2)+f_2(P_1,P_2)\right]\Lambda(P_1)\Lambda(P_2)
\label{eqp},
\end{eqnarray}
where
\begin{eqnarray}
f_1(P_1,P_2)&=&(|\delta|^2-|\gamma|^2)(A^*A-B^*B),
\label{f1}\\
f_2(P_1,P_2)&=&4\textrm{Re}(\delta\gamma^*AB^*).
\label{f2}
\end{eqnarray}
Here the operator
$\sum_{\mathbf{k}}\alpha_\mathbf{k}^+\alpha_\mathbf{k}$
($\sum_{\mathbf{k}}\beta_\mathbf{k}^+\beta_\mathbf{k}$) in $A$ and
$B$ has been replaced by integer $P_1$ ($P_2$) of its
corresponding eigenvalues. The two conditional partition functions
are defined as
\begin{eqnarray}
\Lambda(P_1)=\textrm{tr}_{B(P_1)}e^{-\sum_{\mathbf{k}}\omega_\mathbf{k}\alpha_\mathbf{k}^+\alpha_\mathbf{k}/T},\\
\Lambda(P_2)=\textrm{tr}_{B(P_2)}e^{-\sum_{\mathbf{k}}\omega_\mathbf{k}\beta_\mathbf{k}^+\beta_\mathbf{k}/T},
\end{eqnarray}
where only the states with eigenvalue $P_1$ ($P_2$) of
the operator
$\sum_{\mathbf{k}}\alpha_\mathbf{k}^+\alpha_\mathbf{k}$
($\sum_{\mathbf{k}}\beta_\mathbf{k}^+\beta_\mathbf{k}$) are traced
over. Because of the restriction in the trace, 
the evaluation of $\Lambda(P)$ is a little bit involved.
To proceed, we define a generating
function $G(\lambda)$ for $\Lambda(P)$ in the following manner \cite{xu}. For
any real number $\lambda$ ($|\lambda|\leq1$), we define
\begin{eqnarray}
G(\lambda)=\sum_{P=0}^\infty
\lambda^P\textrm{tr}_{B(P)}e^{-\sum_{\mathbf{k}}\omega_\mathbf{k}\alpha_\mathbf{k}^+\alpha_\mathbf{k}/T}. \label{eq9}
\end{eqnarray}
The Taylor polynomial expansion of function
$G(\lambda)$ at $\lambda=0$ can be written as 
\begin{eqnarray}
G(\lambda)=\sum_{P=0}^\infty \lambda^P\frac{G^{(P)}(0)}{P!},
\label{expansionG}
\end{eqnarray}
where $G^{(P)}(0)$ represents
$G^{(P)}(\lambda)|_{\lambda=0}=d^PG/d\lambda^P|_{\lambda=0}$,
i.e., the derivative of order $P$ of the the function
 $G(\lambda)$ at $\lambda=0$. 
From Eqs.~(\ref{eq9}) and (\ref{expansionG}), 
the coefficient of $\lambda^P$ in the expansion of
$G(\lambda)$ 
is $\Lambda(P)=\textrm{tr}_{B(P)}e^{-\sum_{\mathbf{k}}\omega_\mathbf{k}\alpha_\mathbf{k}^+\alpha_\mathbf{k}/T}$.
Therefore, we have
\begin{eqnarray}
\Lambda(P)=\frac{G^{(P)}(0)}{P!}.\label{eq11}
\end{eqnarray}
It is easy to directly evaluate  Eq.~(\ref{eq9}) to obtain
\begin{eqnarray}
G(\lambda)&=&\prod_{\mathbf{k}}\frac{1}{1-\lambda e^{-\omega_\mathbf{k}/T}}\nonumber\\
&=&e^{-\Omega\int_0^\infty \ln \left ( 1-\lambda e^{-x}
\right)x^2dx} .
\end{eqnarray}
Using the expansion of
\begin{eqnarray}
\ln(1-\xi)=-\xi-\frac{1}{2}\xi^2-\frac{1}{3}\xi^3\cdots
(\textmd{for} |\xi|<1),
\end{eqnarray}
we obtain
\begin{eqnarray}
G(\lambda)=e^{2\Omega\sum_{n=1}^\infty\frac{1}{n^4}\lambda^n}.
\end{eqnarray}
It can be shown that
\begin{eqnarray}
G^{(0)}(0)&=&1,\\
G^{(1)}(0)&=&2\Omega G^{(0)}(0),\\
G^{(2)}(0)&=&2\Omega
\left[\frac{1}{2^3}G^{(0)}(0)+G^{(1)}(0)\right],\\
\vdots\nonumber\\
G^{(P)}(0)&=&2\Omega
(P-1)!\sum_{i=0}^{P-1}\frac{1}{i!(P-i)^3}G^{(i)}(0).
\end{eqnarray}
Using these recursion relations and Eqs.~(\ref{eqp}) and
(\ref{eq11}), we can then evaluate the expectation value of
$S_0^z$. In the same way, the reduced density matrix can be
written as
\begin{eqnarray}
\rho_S(t)&=&\left(\begin{array}{cc}
\rho_{11}(t) & \rho_{12}(t) \\
\rho_{21}(t) & \rho_{22}(t)
\end{array}\right),
\end{eqnarray}
where
\begin{eqnarray}
\rho_{11}(t)&=&\frac{1}{Z}\sum_{P_1=0}^\infty\sum_{P_2=0}^\infty
\left(|\delta|^2AA^*+\delta\gamma^*
AB^*+\delta^*\gamma
A^*B+|\gamma|^2BB^*\right)\Lambda(P_1)\Lambda(P_2),
\label{rho11}\\
\rho_{12}(t)&=&\frac{1}{Z}\sum_{P_1=0}^\infty\sum_{P_2=0}^\infty
\left(|\delta|^2AB^*+\delta\gamma^*
AA+\delta^*\gamma BB^*+|\gamma|^2AB\right)\Lambda(P_1)\Lambda(P_2),
\label{eqpp}\\
\rho_{21}(t)&=&\rho_{12}^*(t),\\
 \rho_{22}(t)&=&\frac{1}{Z}\sum_{P_1=0}^\infty\sum_{P_2=0}^\infty
\left(|\delta|^2BB^*+\delta\gamma^*AB+\delta^*\gamma
A^*B^*+|\gamma|^2AA^*\right)\Lambda(P_1)\Lambda(P_2).
\label{rho22}
\end{eqnarray}

\section{Results and Discussion}
\label{sec:results}

\subsection{Spin inversion and Rabi oscillation decay}

\begin{figure}[tbp]
\includegraphics [width=\linewidth] {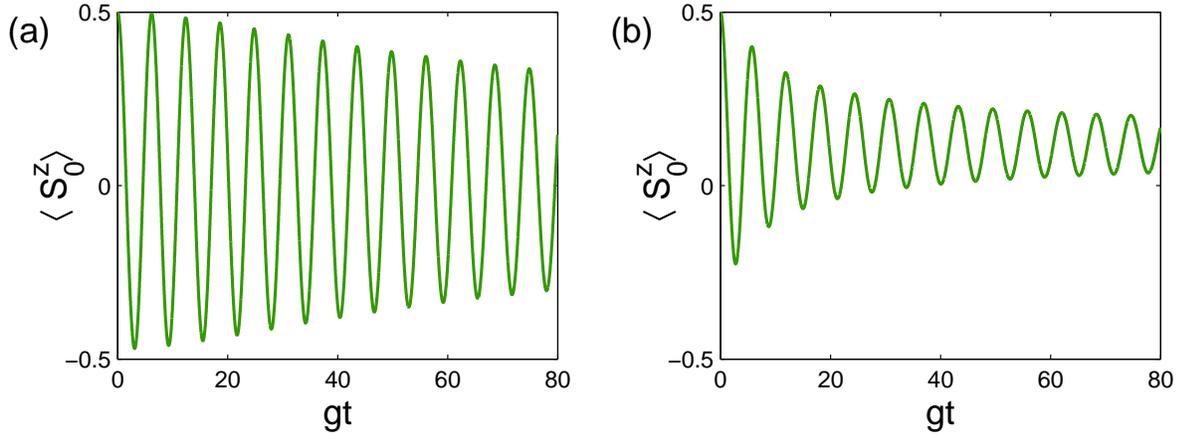}
\caption{Time evolution of $\langle S_0^z\rangle$ 
for different values of (a) $\Omega=2$ and (b) $\Omega=30$.
The initial
qubit state is $|\psi(0)\rangle=|1\rangle$ and other parameters are
$\varepsilon=0$ and
$J_0=0.05g$.}
\label{fig1}
\end{figure}

\begin{figure}[tbp]
\includegraphics [width=\linewidth] {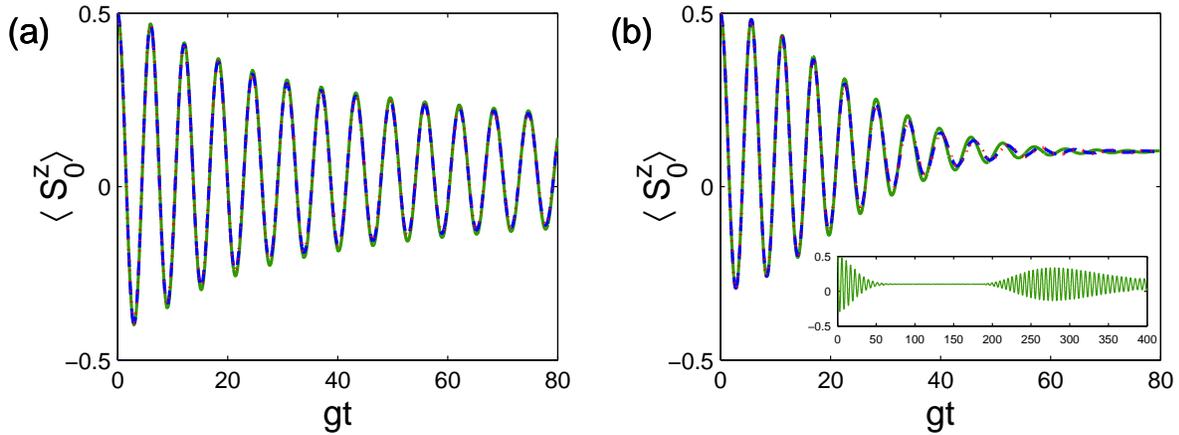}
\caption{(a) Time evolution of $\langle S_0^z\rangle$ 
for different values of
$\Omega=2$, $J_0=0.1g$ (green solid curve), $\Omega=2\times 10$,   
$J_0=0.1g/\sqrt{10}$ (blue dashed curve) and $\Omega=2\times 100$,   
$J_0=0.1g/\sqrt{100}$ (red dotted curve).
The initial
qubit state is $|\psi(0)\rangle=|1\rangle$ and the other parameter is
$\varepsilon=0$. 
(b) Time evolution of $\langle S_0^z\rangle$ 
for different values of
$\Omega=1$, $J_0=0.05g$ (green solid curve), $\Omega=1\times 10$,   
$J_0=0.05g/\sqrt{10}$ (blue dashed curve) and $\Omega=1\times 100$,   
$J_0=0.05g/\sqrt{100}$ (red dotted curve).
The initial
qubit state is $|\psi(0)\rangle=|1\rangle$ and the other parameter is
$\varepsilon=0.5g$. The inset shows the time evolution of the case of
$\Omega=1$ and $J_0=0.05g$ in a larger time interval.
}
\label{fig2}
\end{figure}

In this section, we present and discuss the results we obtain.
We first study the dynamics of the central spin inversion.
Figures \ref{fig1}(a) and \ref{fig1}(b) show the time evolution of $\langle
S_0^z\rangle$ for different values of $\Omega$
in the case of
resonance (i.e., detuning $\varepsilon=0$).
Figure \ref{fig2}(a) shows the time evolution of $\langle
S_0^z\rangle$ also in  
resonance (i.e., detuning $\varepsilon=0$)
but with a different value of 
the system-environment coupling strength $J_0$  
from that of Fig. \ref{fig1}(a). 
For the parameters chosen in
Figs.~\ref{fig1}(a), \ref{fig1}(b) and \ref{fig2}(a),
the driving strength $g$ is much larger than the coupling strength, i.e., 
$g\gg J_0$. As a result, the self-Hamiltonian is dominant over
the interaction with the environment. The
eigenstates of the self-Hamiltonian are
$|+\rangle=(|1\rangle+|0\rangle)/\sqrt{2}$ and
$|-\rangle=(|1\rangle-|0\rangle)/\sqrt{2}$, separated by a large
Rabi frequency $g$. The main influence of the environment on the
central spin is to destroy the initial phase relation between the
states $|+\rangle$ and $|-\rangle$. This leads to the decay of
$\langle S_0^z\rangle$, i.e., Rabi oscillation decay.
From Figs.~\ref{fig1}(a) and \ref{fig2}(a), we can see that 
as expected, increasing the value of $J_0$ results in the increase of
the decay rate of the Rabi oscillations.
Apart from the coupling constant $J_0$,
the important factor $\Omega$, Eq.~(\ref{eq:Omega}), also reflects the
influence of the environment on the central spin as shown in
Fig.~\ref{fig1}(b).  
We can see from Figs.~\ref{fig1}(a), \ref{fig1}(b) and \ref{fig2}(a)
that increasing the
factor $\Omega$ and increasing the coupling constant $J_0$ have similar
effects.
One can observe
that the larger the value of $\Omega$ is, the stronger the decay
of the amplitude of the Rabi oscillations will be. 
As in the case of
the spin-boson model discussed in Ref.~\cite{Irish},  the central
spin inversion does not oscillate around the value of 
$\langle S_0^z\rangle=0$. Its oscillations are, however, 
biased (shifted) a little bit toward the positive value of the initial
$\langle S_0^z(0)\rangle$. With the increase of the value of
$\Omega$ or $J_0$, this effect is enhanced.

It was shown in Ref.~\cite{Dobrovitski} that the form
and the rate of Rabi oscillation decay are useful in experimentally
determining the intrabath coupling strength for a broad class of
solid-state systems \cite{Dobrovitski}. This is also the case in
our problem. The central spin is a two-level system. However, due
to its interaction with the environment, the time evolution of the
central spin inversion consists of different frequencies involved
in the time series of Eqs.~(\ref{eqp})--(\ref{f2}) as well
as  Eqs.~(\ref{coeff_A})--(\ref{eq6}).
The frequencies from Eq.~(\ref{eq6}) can be written as
\begin{equation}
\kappa=\sqrt{\left[J_0(P_2-P_1)-\varepsilon\right]^2+g^2}
\label{kappa_freq}
\end{equation}
for a pair of integers $P_1$ and $P_2$. 
The probability distribution of the frequencies is then
\begin{eqnarray}
\sigma(\kappa)=\frac{1}{Z}\sum_{P=0}^\infty\Lambda(P+n)\Lambda(P),
\end{eqnarray}
where $n=|P_2-P_1|$. It is possible that other pairs of integers
$P'_1$ and $P'_2$ may exist which correspond to the same frequency
$\kappa$ but with $m=|P'_2-P'_1|\neq n$. In such case, we should add
an additional probability
\begin{eqnarray}
\sigma(\kappa)=\frac{1}{Z}\sum_{P=0}^\infty\Lambda(P+m)\Lambda(P).
\label{freq_rel}
\end{eqnarray}
For example, the frequencies of Eq.~(\ref{kappa_freq}) can be rewritten as 
\begin{equation}
\kappa=\sqrt{J_0\left[(P_2-P_1)-(\varepsilon/J_0)\right]^2+g^2}.
\end{equation}
If $n=|P_2-P_1|$ is chosen to be $3$,
then other pairs of integers
$P'_1$ and $P'_2$ with $m=|P'_2-P'_1|=17$  
correspond to the same frequency
$\kappa$ for the parameters  $\varepsilon=0.5g$ and 
$J_0=0.05g$ used in Fig.~\ref{fig2}(b). 
Thus, summing the probability distributions with all
possible combinations of $P_1$ and $P_2$ that correspond to the same frequency
$\kappa$ leads to the final frequency probability distribution. 

\begin{figure}[tbp]
\includegraphics [width=\linewidth] {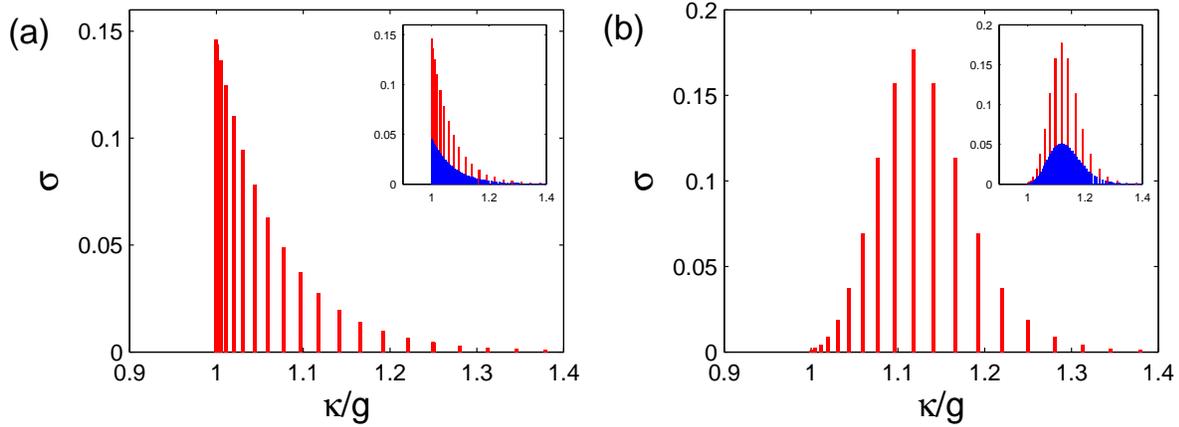}
\caption{(a) Probability  distribution of frequencies for
$\varepsilon=0$, $\Omega=5$, and $J_0=0.05g$. The inset shows the
comparison between the probability distribution of frequency and that
of the case of
increasing $N$ $10$ times (i.e., $\Omega=5\times 10$,   
$J_0=0.05g/\sqrt{10}$) with other parameters unchanged.
(b) Probability  distribution of frequencies for the
parameters used in Fig.~\ref{fig2}(b). The inset shows the
comparison between the probability distribution of frequencies and that
of the case of
increasing $N$ 10 times (i.e., $\Omega=1\times 10$,   
$J_0=0.05g/\sqrt{10}$) with other parameters unchanged.}
\label{fig3}
\end{figure}

Figure \ref{fig3}(a) shows the
frequency probability distribution for the case of 
$\varepsilon=0$, $J_0=0.05g$, and $\Omega=5$. 
It is obvious that the main frequencies are located
near $g$ which is approximately the Rabi frequency.
This is also the case for Figs.~\ref{fig1}(a), \ref{fig1}(b) and
\ref{fig2}(a) where the detuning $\varepsilon=0$. 
The contribution of many different frequencies, 
a consequence of interacting with the AF bath, results in
the Rabi oscillation decay of the central spin. 
For the case of non-zero
detuning $\varepsilon=0.5g$, Fig.~\ref{fig2}(b), with 
$J_0=0.05g$ and $\Omega=1$,
shows a decay of oscillations with an approximated Rabi 
frequency of
$\sqrt{\varepsilon^2+g^2}\approx 1.12g$.
The oscillation residual
amplitude approaches almost zero for a time period of about $65\leq
gt\leq 80$ [see also the inset of Fig.~\ref{fig2}(b)]. 
In such case, different frequencies interfere with each
other and produce the zero amplitude variation period, a behavior
that is called collapse in quantum optics \cite{Scully5}. At
longer times, the oscillation amplitude revives. This is shown
clearly in the inset of Fig.~\ref{fig2}(b). 
Reference \cite{Hanson} reported an experimental observation of a similar
collapse and revival phenomenon for driven spin oscillations,
though the spin bath and the corresponding interactions are
different from those discussed here.

The environmental conditions
affect the dynamics of the central spin and the completeness of
the collapse.
Figure \ref{fig3}(b) shows the frequency probability 
distribution for the central spin inversion
with parameters used in Fig.~\ref{fig2}(b). 
The distribution has a
left-hand-side cutoff at $\kappa_{min}=g$ and 
the center of the distribution is located at
$\kappa_c=\sqrt{\varepsilon^2+g^2}$. 
One can easily understand this from the
frequency relation Eq.~(\ref{kappa_freq}).
It is also obvious  from Eq.~(\ref{kappa_freq}) that the
center of the distribution shifts toward larger frequencies with
the increase of the detuning [this can also be seen from
the comparison between Figs.~\ref{fig3}(a) and  \ref{fig3}(b)]. 
If the detuning $\varepsilon$ exists, it is
possible that the shape of the frequency probability distribution 
becomes approximately
Gaussian, which then results in well-defined collapse and revival
behavior regions \cite{Irish}. If the detuning is zero, only a half side of
the distribution exists [see Fig.~\ref{fig3}(a)] and the central spin
inversion will never show the (complete) collapse and revival behaviors. 
The frequency probability 
distribution is determined also by the coupling strength $J_0$ and the
important factor $\Omega$. With the increase of $J_0$ or $\Omega$, the width
of frequency distribution increases and the probability decreases.
As a result, the decay of the Rabi oscillations is enhanced.

We note that 
the results presented here 
depend, of course, on the number of environment atoms $N$ in each   
sublattice. The   
influence of the number of  atoms $N$ in each   
sublattice on the dynamics of the   
central spin comes from two respects. One is the scaled coupling   
constant $J_0=J'_0//\sqrt{N}$
which is proportional to $1/\sqrt{N}$. The other is the   
factor $\Omega$, Eq.~(\ref{eq:Omega}), which is proportional to $N$. 
We plot in Fig.~\ref{fig2}(a) and Fig.~\ref{fig2}(b)
the time evolutions of $\langle S_0^z\rangle$ in (blue) dashed
curve and (red) dotted curve corresponding, respectively, to 
increasing $N$ by 10 and 100 
times but keeping other parameters unchanged. 
The results show that 
the two curves in dashed and dotted lines almost coincide with each other.
If we increase $N$ by 200 times, then the resultant evolution and that
of increasing $N$ by 100 times become indistinguishable and 
converge toward the same evolution. 
In other words, a nontrivial finite limit exists 
in the thermodynamic limit of $N\rightarrow\infty$. 
This has been shown analytically 
in Refs.~\cite{Yuanepl,YuanNJP,Yuan10} for a spin coupled to an AF
environment without an external {\it ac} driving field.  
We show numerically here that this $N\rightarrow\infty$ limit also
exists in the driven spin case.
One can also notice that 
the difference between the original curve (in solid line) and the two
curves (in dashed and dotted lines, respectively) in Fig.~\ref{fig2} is very   
small in the all time interval shown. This difference in
Fig.~\ref{fig2}(b) with $\varepsilon\neq 0$ is also small but is slightly
larger at large times. 
These indicate that the results presented  
here are very close to the results in the thermodynamic limit.
If we also increase the AF atom number $N$ in $J_0$ and $\Omega$
simultaneously but keep other parameters unchanged, then we can find that
the number of the contributed frequencies in the probability
distribution of frequencies
becomes larger (i.e., the separation between two adjacent
distributed frequencies
becomes smaller or denser) and the value of the distribution becomes
smaller [see the insets in
Figs.~\ref{fig3}(a) and \ref{fig3}(b)]. 
But the width and the shape of the frequency
probability distribution remain almost the same as before [see the insets in
Figs.~\ref{fig3}(a) and \ref{fig3}(b)]. 
The fact that   
the frequency distribution does not change much 
results in the very similar dynamics of   
the driven spin (see Fig.~\ref{fig2}) when we increase $N$.
The existence of nontrivial finite limits as $N\rightarrow\infty$ and
the small difference between the original dynamical evolutions with their
 counterparts in the thermodynamic limit are true for all the dynamic
variables ($\langle S_0^z \rangle$ and von Neumann entropy) presented in this article.

\subsection{von Neumann entropy}

\begin{figure}[tbp]
\includegraphics [width=\linewidth] {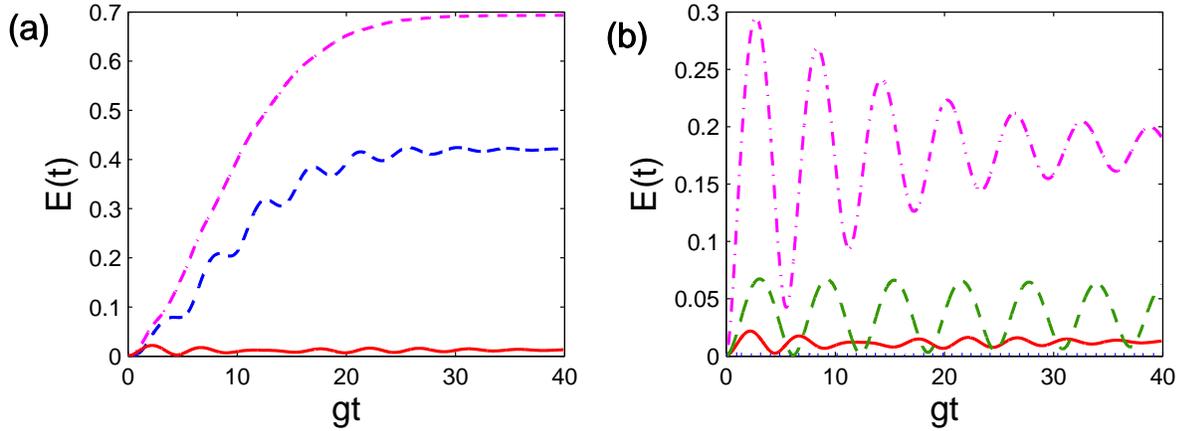}
\caption{
(a) Entropy $E(t)$ for different initial qubit states:
$|\psi(0)\rangle=(|\varphi_1\rangle+|\varphi_2\rangle)/\sqrt{2}$ 
(pink dot-dashed curve), $|\psi(0)\rangle=|1\rangle$ (blue dashed curve),
$|\psi(0)\rangle=|\varphi_1\rangle$ or
$|\psi(0)\rangle=|\varphi_2\rangle$ (red solid curve).
Other parameters are $\varepsilon=g$, $J_0=0.01g$, $\Omega=20$.
(b) Entropy $E(t)$ for 
(i) pink dot-dashed curve: $J_0=0.03g$, $\varepsilon=0$, 
(ii) green dashed curve: $J_0=0.01g$, $\varepsilon=0$, 
(iii) red solid curve: $J_0=0.01g$, $\varepsilon=g$,
and (iv) blue dotted curve: $J_0=0.01g$, $\varepsilon=3g$. 
 The qubit initial state is $|\psi(0)\rangle=|\varphi_1\rangle$ or
$|\psi(0)\rangle=|\varphi_2\rangle$, 
and other parameter is  $\Omega=20$.
}
\label{fig5}
\end{figure}

The purity of a mixed state can be  measured by the von Neumann entropy
\begin{eqnarray}
E(t)=-\textrm{tr}_S[\rho_S(t)\ln\rho_S(t)].
\end{eqnarray}
After diagonalizing the reduced density matrix $\rho_S(t)$, we obtain
the von Neumann entropy
\begin{eqnarray}
E(t)=-\sum_{k=1}^2p_k(t)\ln p_k(t),
\end{eqnarray}
where $p_k(t)$ are the eigenvalues of the reduced density matrix
$\rho_S(t)$ and can be expressed as
\begin{eqnarray}
p_{1,2}(t)=\frac{\rho_{11}(t)+\rho_{22}(t)\pm\sqrt{[\rho_{11}(t)-\rho_{22}(t)]^2+4\rho_{12}(t)\rho_{21}(t)}}{2}.
\end{eqnarray}
We investigate the von Neumann entropy of the qubit system in
Fig.~\ref{fig5}.
In Fig.~\ref{fig5}(a), for the initial state $|\psi(0)\rangle=|1\rangle$, the
entropy $E(t)$ oscillatorily approaches its maximal value of 0.4228
(in blue dashed curve). If
the initial state is one of the eigenstates of the qubit
self-Hamiltonian, i.e., $|\varphi_1\rangle$ or
$|\varphi_2\rangle$, the von Neumann entropy $E(t)$ remains small
with time and
approaches a very small value in the long-time limit (in red solid line).
In the opposite, if we choose the superposition state
$|\psi(0)\rangle=(|\varphi_1\rangle+|\varphi_2\rangle)/\sqrt{2}$ 
as the initial state, results show that the von Neumann entropy $E(t)$
increases monotonously
to the maximal value of $\ln2=0.6971$ (in pink dot-dashed line), 
representing a completely
mixed state. That is to say, this initial state gets maximally entangled
with the environment in the course of the evolution.  
This initial superposition state is thus frangible to the influence of the
environment and should be avoided being used in the quantum information
processing in the presence of the AF environment. 
In all the cases, the entropy $E(t)$ increases with the
increase of the factor $\Omega$ and the coupling constant strength $J_0$.

We discuss next the pointer states of the qubit from the
perspective of the von Neumann entropy \cite{Paz0,Zurek93,Zurek03}.
For $g=0$, i.e., no driving field, it is
obvious that the eigenstates of the self-Hamiltonian, $\mu_0
S_0^z$, are the pointer states as the self-Hamiltonian commutes with
the interaction Hamiltonian with the environment.
In this case, a generic quantum state decays, after the decoherence time, 
into a mixture of pointer states. 
Thus the decoherence behavior can be described effectively by
the off-diagonal elements (decoherence factor)
of the reduced density matrix in the pointer state
basis of the system. This is exactly the cases studied in 
Refs.~\cite{Yuanepl,YuanNJP},
where the off-diagonal elements (decoherence factor) vanish in the
long-time limit in the eigenstate basis of the self-Hamiltonian 
or of the interaction Hamiltonian with the environment regardless of
the different initial states.
If the driving field strength $g\neq0$, 
it is difficult to identify the exact pointer states of the
system (especially in the non-perturbative and non-Markovian regime)
as the self-Hamiltonian does not commute with
the interaction Hamiltonian with the bath. 
Pointer states may be defined as the states which become minimally
entangled with the environment in the course of their 
evolution \cite{Paz0,Zurek93}. 
An operational definition in terms of the von Neumann entropy, 
introduced in Refs.~\cite{Paz0,Zurek93,Zurek03}, 
is that pointer states are obtained by minimizing 
the von Neumann entropy over the initial
state $|\psi(0)\rangle$ and requiring that the answer be robust when
varying the time $t$. 
In general situations, 
pointer states result from the interplay between self-evolution and
interaction with the environment, and thus their dynamical selection by
the environment are complicated.
For $g>>J_0$ or/and $\varepsilon>>J_0$, the pointer states turn
out (approximately) to be the eigenstates of the self-Hamiltonian 
\cite{Cucchietti,Paz0}.
Here the eigenstates $|\varphi_1\rangle$ and $|\varphi_2\rangle$ of the self-Hamiltonian $H_S=\varepsilon S_0^z+g
S_0^x$ defined in Eq.~(\ref{Hs_rot})
can be written as
\begin{eqnarray}
\left(\begin{array}{c}
|\varphi_1\rangle\\
|\varphi_2\rangle
\end{array}\right)
&=& U 
\left(\begin{array}{c}
|1\rangle\\
|0\rangle
\end{array}\right)\\
&=&\left(\begin{array}{cc}
U_{11} & U_{12} \\
U_{21} & U_{22}
\end{array}\right)
\left(\begin{array}{c}
|1\rangle\\
|0\rangle
\end{array}\right),
\label{U_transform}
\end{eqnarray}
where
\begin{eqnarray}
U_{11}&=&\frac{1}{\sqrt{2}}\frac{g}{\sqrt{\varepsilon^2+g^2-\varepsilon\sqrt{\varepsilon^2+g^2}}},\\
U_{12}&=&\frac{1}{\sqrt{2}}\frac{\sqrt{\varepsilon^2+g^2}-\varepsilon}{\sqrt{\varepsilon^2+g^2-\varepsilon\sqrt{\varepsilon^2+g^2}}},\\
U_{21}&=&\frac{1}{\sqrt{2}}\frac{g}{\sqrt{\varepsilon^2+g^2+\varepsilon\sqrt{\varepsilon^2+g^2}}},\\
U_{22}&=&-\frac{1}{\sqrt{2}}\frac{\sqrt{\varepsilon^2+g^2}+\varepsilon}{\sqrt{\varepsilon^2+g^2+\varepsilon\sqrt{\varepsilon^2+g^2}}}.
\label{U_22}
\end{eqnarray}
Thus for the Hamiltonian parameters chosen in Fig.~\ref{fig5}(a) which
is in
a regime where the self-Hamiltonian of the system dominates, 
it is expected
that the pointer states are very close to the 
eigenstates $|\varphi_1\rangle$ and $|\varphi_2\rangle$ \cite{Cucchietti,Paz0}. 
By changing different initial states near the
eigenstates $|\varphi_1\rangle$ or $|\varphi_2\rangle$ 
with other parameters fixed, it is found that 
the eigenstate $|\varphi_1\rangle$ or $|\varphi_2\rangle$ 
produces the minimum entropy increase with time (the red solid line).
This suggests that the
eigenstates $|\varphi_1\rangle$ and $|\varphi_2\rangle$
are dynamically selected by the environment as the preferred 
pointer states in the self-Hamiltonian dominant regime.  
Moreover, when the energy parameters $g$ and $\varepsilon$ 
of the self-Hamiltonian of the central spin system
are increased further with respect to the system-bath coupling
strength $J_0$
(i.e., the self-Hamiltonian dominates even more),
the values of entropy evolution are lower 
(the green dashed curve is lower than the pink dot-dashed curve in the
zero-detuning case, and the blue dotted curve 
that is close to zero is lower than the red
solid curve in the detuning case) as shown  
in Fig.~\ref{fig5}(b). 
We have also checked that the four curves 
with initial state $|\varphi_1\rangle$ or $|\varphi_2\rangle$ 
in Fig.~\ref{fig5}(b) are, respectively,
the minimum entropy curves 
in the course of evolution among 
those entropy evolution curves with initial states varied near the eigenstate 
of $|\varphi_1\rangle$ or $|\varphi_2\rangle$ but with other
parameters fixed.
This confirms furthermore again that in the regime where the self-Hamiltonian
of the system dominates (with or without detuning), the eigenstates of
the system self-Hamiltonian emerge as the preferred pointer states.

\section{Conclusions} 
\label{sec:conclusions}

We investigate the decoherence of a qubit
in an AF environment, in the presence of a driving
field. The difficulty of our problem lies in the fact that the
self-Hamiltonian does not commute with the interaction Hamiltonian
and the internal dynamics (coupling) of the spin bath are taken into
consideration at the same time. The spin-wave approximation is
used to map the spin operators of the AF environment onto bosonic
operators in the low-temperature and low-excitation limit. 
Then the resultant model is solved exactly, even in the case of
multi-environment modes and finite environment temperatures. 
Our approach includes the 
environment dynamics, the qubit dynamics, and the
quantum correlations between them. The influence of the
AF environment on the qubit is found to depend on two
factors, i.e., the coupling constant $J_0$ and a dimensionless
factor $\Omega=\frac{NT^3}{4\sqrt{2}\pi^2M^{3/2}J^3}$. Increasing
the two factors, the decay time of the Rabi oscillations becomes
shorter, the decoherence of the qubit is enhanced, and the qubit
state becomes more mixed with time. 
The time evolution of the Rabi oscillations also depends on the
detuning between the driving frequency and the qubit Larmor frequency.   
If the detuning exists, the Rabi oscillations may
show a behavior of collapses and revivals; however, if the
detuning is zero, such a behavior will not appears.
This can be understood in terms of the weighted frequency distribution
investigated here. 
Also the decoherence and the
pointer states of the qubit are discussed from the perspective
of the von Neumann entropy. It is found that the eigenstates of the qubit
self-Hamiltonian are dynamically selected by the environment as the
preferred pointer states 
in the weak system-environment coupling limit 
(or in the self-Hamiltonian dominant regime).

\ack
H.S.G. would
like to acknowledge support from the National Science
Council, Taiwan, under Grant No. 97-2112-M-002-012-MY3, 
support from the Frontier and Innovative Research Program 
of the National Taiwan University under Grants No. 99R80869 and 
No. 99R80871,
and support from the focus group
program of the National Center for Theoretical Sciences, Taiwan.
H.S.G. is also 
grateful to the National Center for High-performance Computing, Taiwan, 
for computer time and facilities.
X.Z.Y. acknowledges support from the National Science Foundation of   
China under Grant No. 10874117. X.Z.Y. also acknowledges the support by the
National Science Foundation (Grant No. 11091240282) to do research in   
ICTP (SMR2154), and  thanks ICTP and APCTP for their
invitations as a visiting professor and for their hospitality.

\end{document}